\documentclass{article}
\usepackage[T1]{fontenc}
\usepackage[accepted]{icml2026}

\usepackage{booktabs}
\usepackage{graphicx}
\usepackage{url}
\usepackage{microtype}
\usepackage{amsmath}
\usepackage{amssymb}

\usepackage{hyperref}
\usepackage{xurl}

\icmltitlerunning{Probes Surface Code-Security Signals}

\begin{document}
\twocolumn[
\icmltitle{Activation Probes Surface Code-Security Signals\\
that the Model's Output Misses}
\icmlsetsymbol{equal}{*}
\begin{icmlauthorlist}
\icmlauthor{Ivan Wiryadi}{anon}
\end{icmlauthorlist}
\icmlaffiliation{anon}{Independent}
\icmlcorrespondingauthor{Ivan Wiryadi}{inbox.ivn@gmail.com}
\icmlkeywords{AI governance, code review, coding agents, linear probes,
activation monitoring, vulnerability detection}
\vskip 0.3in
]
\printAffiliationsAndNotice{}

\begin{abstract}
AI coding agents now write a growing share of production code, and
human security review does not scale at the rate code is generated.
The agents in widest use are closed-weight, so a deploying team
cannot read their internals. It can instead run an open-weight
model as a reviewer over the agent's output. That reviewer's
activations are readable. We ask whether reading those activations
recovers a security signal that simply asking the same reviewer
misses. We fit a single
linear probe per model on a corpus of paired vulnerable-and-fixed
Python functions, then test it without retraining on real
disclosed vulnerabilities whose weakness type the probe never saw
in training, across five open-weight reviewer models. On the
vulnerabilities fixed by changing a single function, the probe
scores the vulnerable function above its fix on 61--67\% of cases
for every model, beating the 50\% chance line. It also beats the
same model's prompted YES/NO win-rate read from its logits, under
every prompt we try. Asking the model for a written verdict, even with
chain-of-thought, returns the same answer on the vulnerable and
fixed function most of the time and so cannot tell them apart. Model activations carry a
code-security signal that prompting the same model misses. 
\end{abstract}

\section{Introduction}
\label{sec:intro}

AI coding agents, such as \emph{Claude Code}, \emph{Cursor},
\emph{Windsurf}, and \emph{OpenAI Codex}, now author a growing
share of production code~\citep{anthropic2026claudecode,
wang2025aicodewild, rooney2025vibecoding}. Empirical audits find that 24--29\% of
AI-assistant-generated Python and JavaScript snippets
contain at least one security weakness, spanning tens of distinct
weakness types~\citep{fu2025copilotsecurity}. Human security review
does not scale at the rate code is now generated, so some automated
check in the loop is needed to triage what a reviewer looks at.

Current practice offers two such checks, and both are known to
miss vulnerabilities. Static analyzers produce a high
false-positive rate even on hand-crafted
micro-benchmarks~\citep{dubniczky2025castle}, and prompting a
coding LLM for a binary verdict is non-robust: trivial rewrites
(renaming a variable, adding an unrelated library call) flip the
answer in 17--26\% of cases for GPT-4 and
PaLM2~\citep{ullah2024secllmholmes}. We confirm a related failure
mode on paired vulnerable-and-fixed files: the model emits the
\emph{same} verdict on both most of the time, so its stated answer
cannot tell them apart.

Linear probes on residual-stream activations encode
safety-relevant signals that a model's own output can
hide~\citep{mckenzie2024linear, alain2016probes, orgad2024knowmore,
azaria2023internal, marks2024geometry}. Similarly, \textsc{LPASS}
applied linear probes on compressed-LLM activations to
vulnerability classification on C/C++, training and testing on the
same weakness types~\citep{lpass2025linear}. We extend this line of
work along the axes that matter for a reviewer: we ask whether a
linear probe recovers a security signal on Python that the model's
own prompted output misses, and whether it transfers to weakness
types held out of training and to real disclosed vulnerabilities.
The signal is read straight off the model's hidden state. The
longer-term goal is a control a team can self-host: an open-weight
reviewer over a coding agent's output that flags changes worth
re-reviewing, without access to the agent's own internals. Whether
the signal can support that is what we begin to test here.

Across five open-weight code LLMs, the probe transfers to these
unseen real-world vulnerabilities and outranks the same model's
prompted output on them, a comparison we make precise in
Section~\ref{sec:method}. Wiring the probe's verdict back into the
coding agent is left to future work.

\section{Method}
\label{sec:method}

\paragraph{Reviewer models.} Five open-weight code LLMs spanning
four architecture families: Qwen2.5-Coder 7B and
14B~\citep{hui2024qwen25coder}, DeepSeek-Coder
33B~\citep{guo2024deepseekcoder},
Devstral-Small-2505~\citep{mistral2025devstral}, and
Llama-3.1-8B~\citep{grattafiori2024llama3}. Each is the instruct
variant.

\paragraph{Data.} We train on SVEN and evaluate on
\textsc{PatchEval}, both labelled by CWE. A CWE (Common Weakness
Enumeration) is a \emph{type} of software flaw, such as CWE-089 SQL
injection; a CVE (Common Vulnerabilities and Exposures) is a single
disclosed vulnerability, which carries one or more CWE labels. For
training we use the four SVEN~\citep{he2023sven} CWEs with
substantial Python coverage (CWE-022, 078, 079, 089), which we call
\textbf{SVEN-Python}; SVEN supplies each CWE as paired before-fix
(vulnerable) and after-fix (secure) functions. For evaluation we
use 234 Python CVEs from \textsc{PatchEval}~\citep{wei2025patcheval}
whose bug type is \emph{unseen}: no CWE in the CVE's full label set
overlaps the four SVEN-training CWEs. (The remaining PatchEval CVEs,
whose labels do overlap training, form a seen reference reported
only in Appendix~\ref{app:seen}.)

\paragraph{Probe fitting.} For each model we extract the residual
stream via \texttt{nnsight}~\citep{fiottokaufman2024nnsight},
hooking the input of \texttt{layers[$L$].input\_layernorm}. We
sweep layers, token-pooling strategies (mean, max, last,
sliding-window-max), and $L_2$ strength, selecting the
configuration that maximises 5-fold \texttt{GroupKFold}
AUC~\citep{pedregosa2011sklearn}. Both
the train/eval split and the cross-validation folds are grouped
by \emph{project}: every repository lands entirely in train or
entirely in eval. This keeps the probe from being scored on code it
already saw in training, whether from the same project or the same
vulnerability. Such duplication and leakage are known to inflate
code-model evaluations~\citep{allamanis2019duplication,
ding2025leakage}. The selected layer and pooling vary by model;
per-model choices are in Appendix~\ref{app:training}.

\paragraph{Out-of-distribution evaluation.} We apply the
SVEN-trained probes zero-shot to the unseen \textsc{PatchEval}
CVEs. For each CVE we score the vulnerable function
($\hat{p}_\text{vul}$) and its post-fix counterpart
($\hat{p}_\text{fix}$) and report the paired \emph{win-rate}: the
fraction of CVEs whose risk gap $\hat{p}_\text{vul} -
\hat{p}_\text{fix}$ is positive, with exact ties dropped. We attach
a Wilson 95\% interval and an
exact-binomial sign test against chance (0.5).
The probe scores one function at a time, so a CVE fixed by changing
a single function compares directly; these single-function CVEs are
our headline slice and are unaffected by the pairing step below.
When a patch touches several functions, we pair each vulnerable
function with its fix by file path and definition name (dropping
any vulnerable function with no matching fix) and average the
matched-pair gaps into one risk gap per CVE, so the CVE stays the
unit of analysis. We report the single-function slice and the full
set separately.

\paragraph{Prompted comparator.} Against the probe we run the
same model's prompted output, varying the prompt three ways and
reading the answer two ways. The prompt asks with no examples,
with a few worked vulnerable/fixed examples from held-out weakness
types (\emph{few-shot}), or with a chain-of-thought trace before
the verdict, generated with vLLM~\citep{kwon2023vllm}. The
\emph{logit} reading takes the renormalised YES-versus-NO
probability at the answer token~\citep{mckenzie2024linear}, a
continuous score comparable to the probe; as for the probe, a CVE
is scored \emph{correct} when this score is higher on the
vulnerable function than on its fix. The \emph{token} reading
instead takes the literal ``yes'' or ``no'' string the model
writes; here the two functions usually receive the same word, so
the pair ties and cannot be ranked. We compare the probe against the
strongest prompting per model with a paired McNemar
test on the same CVEs. The strongest
prompting is that model's highest-win-rate prompt-and-readout;
picking the comparator's best configuration only makes the test
harder for the probe. A tie counts as an incorrect prompting
verdict.

\section{Results}
\label{sec:results}

\begin{figure}[t]
\centering
\includegraphics[width=\linewidth]{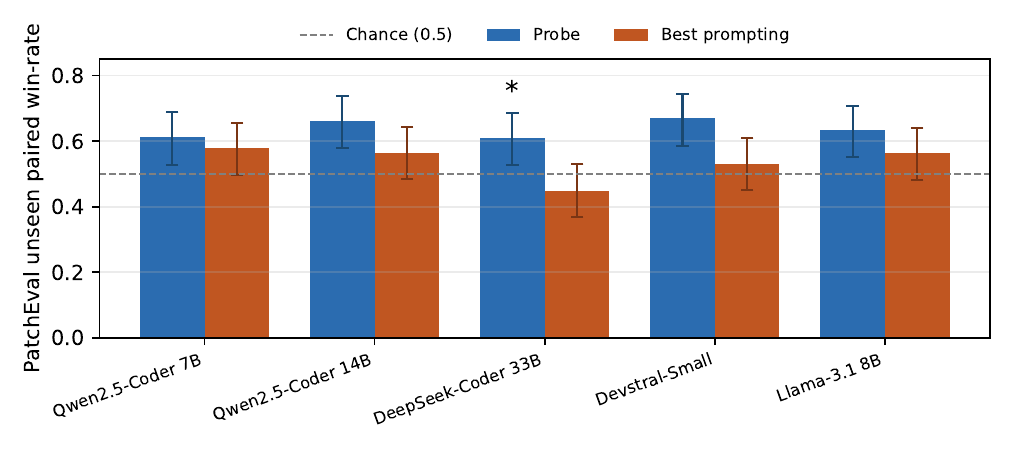}
\caption{Paired win-rate on 147 unseen real-world vulnerabilities:
probe versus the same model's best prompting, per reviewer model.
A star marks a significant probe advantage over that model's
\emph{best} prompt-and-readout (paired McNemar, $p < 0.05$);
Table~\ref{tab:elicitation} stars the advantage over each
individual prompt mode.}
\label{fig:headline}
\end{figure}

\paragraph{The probe scores vulnerable functions higher than
their fixes on unseen bug types.} We evaluate on 234 unseen
real-world vulnerabilities from \textsc{PatchEval}. Of these, 147
are fixed by changing a single function, the cleanest comparison.
There the probe scores the vulnerable function above its fix on
61--67\% of cases
(Figure~\ref{fig:headline}, Table~\ref{tab:patcheval-ood}). This
fraction is the paired \emph{win-rate}: how often the probe ranks
the vulnerable function above its own fix.
The Wilson 95\% interval on it stays above the 50\% chance line for
every model, and the win-rate is above 0.5 by the sign test ($p <
0.05$ for every model). On the full set of 234 the win-rate is
similar (62--69\%).
The effect is modest but holds across all five models and four
architecture families. In-distribution, the same probe separates
vulnerable from fixed functions at a fixed $\tau{=}0.5$
threshold (AUC 0.77--0.85, accuracy 0.72--0.77;
Table~\ref{tab:sven-python}).

\begin{table*}[t]
\centering
\small
\caption{In-distribution detection on SVEN-Python ($n{=}152$). Probe acc@0.5 is the probe's accuracy at a $0.5$ threshold.}
\label{tab:sven-python}
\setlength{\tabcolsep}{3pt}
\begin{tabular}{lcccc}
\toprule
Model & Probe AUC & Probe acc@0.5 & Prompted acc & Probe F1 \\
\midrule
Qwen2.5-Coder 7B & 0.846 & 0.763 & 0.524 & 0.760 \\
Qwen2.5-Coder 14B & 0.821 & 0.737 & 0.518 & 0.733 \\
DeepSeek-Coder 33B & 0.844 & 0.770 & 0.488 & 0.755 \\
Devstral-Small & 0.771 & 0.737 & 0.500 & 0.747 \\
Llama-3.1 8B & 0.787 & 0.717 & 0.500 & 0.703 \\
\bottomrule
\end{tabular}
\end{table*}

\begin{table}[t]
\centering
\footnotesize
\caption{Probe paired win-rate on the unseen-bug-types
\textsc{PatchEval} slice, single-function CVEs ($n{=}147$ pairs; no
CWE overlaps the four SVEN-training CWEs). CI: Wilson 95\%; $p$:
sign test vs.\ 0.5. The seen-bug-types reference is in
Appendix~\ref{app:seen}.}
\label{tab:patcheval-ood}
\setlength{\tabcolsep}{4pt}
\begin{tabular}{@{}lccc@{}}
\toprule
Model & WR (\%) & 95\% CI & $p$ \\
\midrule
Qwen2.5-Coder 7B & 61.2 & 52.8--68.8 & 0.011 \\
Qwen2.5-Coder 14B & 66.2 & 57.8--73.7 & 2.4e-04 \\
DeepSeek-Coder 33B & 60.8 & 52.7--68.5 & 0.012 \\
Devstral-Small & 66.9 & 58.6--74.3 & 9.9e-05 \\
Llama-3.1 8B & 63.4 & 55.2--70.9 & 0.002 \\
\bottomrule
\end{tabular}
\end{table}

\paragraph{The same model's prompted output does not.} We prompt
the same model three ways (no examples, a few examples, and
chain-of-thought) and read its YES/NO answer two ways: from its
logits or from the verdict it writes (Table~\ref{tab:elicitation}).
Reading the answer from the logits, the probe's win-rate is higher
than the model's prompted output for every one of the five models,
under each of the three prompts. By a paired McNemar test the
probe significantly beats no-shot prompting on three of the five
models and few-shot prompting on four; chain-of-thought read from
the logits is significant on only one of the five, and the probe's
win-rate is still higher for every model. We say nothing here
about reading the written verdict, which ties on most pairs and so
ranks too few to compare (next).

\begin{table}[t]
\centering
\footnotesize
\caption{Probe vs.\ prompted win-rate on unseen single-function vulnerabilities. $^{*}$: probe significantly more accurate (paired McNemar, $p{<}0.05$).}
\label{tab:elicitation}
\setlength{\tabcolsep}{2pt}
\begin{tabular}{lcccc}
\toprule
 & Probe & \multicolumn{3}{c}{Prompting (\%)} \\
\cmidrule(lr){2-2} \cmidrule(lr){3-5}
Model & WR (\%) & no-shot & few-shot & CoT \\
\midrule
Qwen2.5-Coder 7B & \textbf{61.2} & 43.8$^{*}$ & 44.5$^{*}$ & 57.8 \\
Qwen2.5-Coder 14B & \textbf{66.2} & 49.0 & 56.5 & 55.5 \\
DeepSeek-Coder 33B & \textbf{60.8} & 33.6$^{*}$ & 35.4$^{*}$ & 44.7$^{*}$ \\
Devstral-Small & \textbf{66.9} & 48.6$^{*}$ & 45.6$^{*}$ & 53.1 \\
Llama-3.1 8B & \textbf{63.4} & 50.3 & 49.3$^{*}$ & 56.2 \\
\bottomrule
\end{tabular}
\end{table}

\paragraph{The logits separate the pair; the written verdict
does not.} The way the answer is read matters more than the
prompt. Reading the YES-versus-NO probability off the logits gives
a rankable score even when it is near chance, because the
probability still differs between the vulnerable and the fixed function.
Reading the verdict the model writes instead collapses: the model
writes the \emph{same} word for both functions on 72--97\% of
cases, so the pair ties and cannot be ranked.\footnote{Scored over
the full slice, the written verdict is \emph{correct} (YES on the
vulnerable function and NO on its fix) on only 1--13\% of pairs; on
the rest the model writes the same verdict for both functions.
Appendix~\ref{app:prompted} (Table~\ref{tab:app-correct-agree})
breaks this down by prompt and readout.} The internal
probability carries a signal the model's own stated answer throws
away, the same gap the activation probe exploits more directly.

\section{Discussion and Future Work}
\label{sec:future}

\paragraph{Future work.} Wiring the probe into a coding agent's
post-write hook, and evaluating it on the code those agents
actually emit, is the natural next step. Our sweep covers mean,
max, last, and sliding-window-max pooling with a logistic probe;
attention-based probes and a probe-architecture comparison remain
untried. Turning the paired signal into a calibrated detector with an
absolute threshold is the main step toward deployment. Extension
beyond Python is also open.

\paragraph{Limitations.} Our out-of-distribution evaluation measures
a paired comparison: whether the probe scores a vulnerable function
above its own fix (win-rate 61--67\%). On unseen weakness types this
is a usable ranking signal. The probe leads every model on win-rate. Its advantage over the strongest prompting
(chain-of-thought from the logits) is narrow and not always
significant; the clear gap is over no-shot and few-shot prompting.
We also worked with limited set data and expanding it is future work. 
Finally, we train only on the four SVEN CWEs
with substantial Python coverage. The C/C++-dominant classes scored
at chance in preliminary runs; whether that reflects a language,
sample-size, or weakness-type effect is open.

\paragraph{LLM usage.} Claude (Anthropic) was used extensively
throughout the project for writing, editing, formatting, and
coding, and for parts of the analysis. The authors set the design
and framing decisions and reviewed outputs. Any remaining errors
are the authors' own.

\bibliographystyle{icml2026}
\bibliography{paper}

\clearpage
\appendix
\section{Appendix}
\label{sec:appendix}

\subsection{Reviewer models and probe training}
\label{app:training}

All five reviewer models are loaded with
HuggingFace Transformers~\citep{wolf2020transformers} in float16, with
no quantisation, on a single 80\,GB H100; running every model at the
same precision avoids a quantisation confound across the comparison.
Activations are extracted with
\texttt{nnsight}~\citep{fiottokaufman2024nnsight} at the input of
\texttt{layers[$L$].input\_layernorm}. Probes are trained in
PyTorch~\citep{paszke2019pytorch}; layer-sweep selection and the
held-out AUC use scikit-learn~\citep{pedregosa2011sklearn};
statistical tests use SciPy~\citep{virtanen2020scipy}.

For each model we sweep the layer, the pooling strategy (mean, max,
last, sliding-window-max), and the $L_2$ constant $C \in
\{10^{-2}, 10^{-1}, 1\}$, choosing the configuration with the
highest 5-fold \texttt{GroupKFold} AUC (Section~\ref{sec:method}).
Table~\ref{tab:app-training} reports the selected configuration,
its cross-validation and held-out AUC on the 152-sample
SVEN-Python eval split, and the same model's prompted YES/NO
accuracy on that split for reference. At a $0.5$ threshold the
probe scores 0.72--0.77 accuracy, while the prompted YES/NO answer
stays near chance (0.49--0.52), a gap of 21--28 points on the
probe's own training distribution.

\begin{table}[!htbp]
\centering
\scriptsize
\setlength{\tabcolsep}{2.5pt}
\caption{Per-model probe setup and held-out SVEN-Python results.
Pooling: max, or sliding-window-max with window~16 (swim). All
models use $C{=}10^{-2}$. The prompted column is the same model's
YES/NO accuracy on the SVEN-Python eval split.}
\label{tab:app-training}
\begin{tabular}{lccccc}
\toprule
Model & Layer & Pool & AUC (CV) & AUC (eval) & Prompted \\
\midrule
Qwen2.5-7B    & 11 & max  & 0.841 & 0.846 & 0.524 \\
Qwen2.5-14B   & 21 & swim & 0.838 & 0.821 & 0.518 \\
DeepSeek-33B  &  7 & max  & 0.831 & 0.844 & 0.488 \\
Devstral      & 21 & max  & 0.846 & 0.771 & 0.500 \\
Llama-3.1-8B  & 17 & max  & 0.846 & 0.787 & 0.500 \\
\bottomrule
\end{tabular}
\end{table}

\subsection{Seen-bug-types reference}
\label{app:seen}

The 169 \textsc{PatchEval} CVEs whose CWE label set overlaps the
four SVEN-training CWEs form a \emph{seen} reference (the main text
reports only the 234 unseen CVEs). This seen slice is the full seen
set (single- and multi-function CVEs), whereas the main-text
headline is the single-function unseen slice, so the two are not a
strict like-for-like comparison. On the seen slice the probe's
paired win-rate is higher (Table~\ref{tab:app-seen}), as expected
when the weakness type was in training; the unseen result in the
main text is the generalisation claim.

\begin{table}[!htbp]
\centering
\scriptsize
\setlength{\tabcolsep}{3.5pt}
\caption{Probe paired win-rate on the seen-bug-types
\textsc{PatchEval} slice ($n{=}169$ CVE pairs whose labels overlap
the four SVEN-training CWEs). CI: Wilson 95\%; $p$: sign test vs.\
0.5.}
\label{tab:app-seen}
\begin{tabular}{lccc}
\toprule
Model & WR (\%) & 95\% CI & $p$ \\
\midrule
Qwen2.5-Coder 7B   & 75.9 & 68.8--81.9 & 2.5e-11 \\
Qwen2.5-Coder 14B  & 79.8 & 72.8--85.3 & 2.2e-14 \\
DeepSeek-Coder 33B & 64.5 & 57.0--71.3 & 2.0e-04 \\
Devstral-Small     & 67.1 & 59.4--73.9 & 2.1e-05 \\
Llama-3.1 8B       & 66.3 & 58.8--73.0 & 3.4e-05 \\
\bottomrule
\end{tabular}
\end{table}

\paragraph{Full unseen set.} On the full set of 234 unseen CVEs
(single- and multi-function), the probe's paired win-rate is 66.4\%
(Qwen2.5-Coder 7B), 69.4\% (14B), 61.6\% (DeepSeek-Coder 33B),
64.7\% (Devstral-Small), and 66.7\% (Llama-3.1 8B), all above 0.5 by
the sign test ($p < 0.001$ for every model). This is the 62--69\%
range quoted in Section~\ref{sec:results}.

\subsection{Prompted comparator: \emph{correct} vs \emph{agree}
on PatchEval}
\label{app:prompted}

The prompt asks the model, as a security code reviewer, whether the
function contains a vulnerability and to answer YES or NO; the
few-shot prefix adds worked examples over weakness types held out of
the eval set; the chain-of-thought variant asks for step-by-step
reasoning ending in an explicit verdict. Out of the five models
tested, four comply with the instruction, while DeepSeek-Coder 33B
tends to refuse or fail to follow it more often (no-verdict rate
16\% no-shot, 9\% few-shot, 2\% CoT), so its prompted win-rates are
on a smaller compliant set.

For the prompted comparator we report the two paired quantities the
main text turns on, on the unseen single-function
\textsc{PatchEval} slice. A pair is \emph{correct} when the model
ranks the vulnerable function above its fix, and the model
\emph{agrees} with itself when it assigns the \emph{same} verdict
to both functions (a tie). The two readouts behave oppositely. The
continuous \emph{logit} readout almost never ties (agree
$\le$1\%), so its \emph{correct} rate is measured over the full
slice, and it sits at or below chance (34--58\%). The binary
\emph{token} readout ties on most pairs (agree 72--97\%), so its
\emph{correct} rate is computed over only the few non-tied
remainder and is not comparable. The probe avoids this failure: it
produces a continuous score on every pair, so there is always
something to rank.

\begin{table}[!htbp]
\centering
\scriptsize
\setlength{\tabcolsep}{3.5pt}
\caption{Prompted comparator on the unseen single-function slice,
as paired \emph{correct} (ranks vulnerable above fix) and
\emph{agree} (same verdict on both) rates. Range across the five
models. The \emph{token} \emph{correct} column is over the small
non-tied remainder only.}
\label{tab:app-correct-agree}
\begin{tabular}{lcc}
\toprule
Prompt $\times$ readout & Correct (\%) & Agree / tie (\%) \\
\midrule
no-shot, logit   & 34--50 & $\le$1 \\
few-shot, logit  & 35--57 & $\le$1 \\
CoT, logit       & 45--58 & 0 \\
\midrule
no-shot, token     & 36--75 & 90--97 \\
few-shot, token    & 10--85 & 86--93 \\
CoT, token         & 40--65 & 72--86 \\
\bottomrule
\end{tabular}
\end{table}

\end{document}